\documentclass[doublecol]{epl2} 
\usepackage{amssymb}
\usepackage{amsmath}

\title{Interaction of hopfions of charge 1 and 2 from product ansatz}

\author{A.~Acus\inst{1,2}, E.~Norvai\v{s}as\inst{1} and Ya. Shnir\inst{3,4,5}}
\shortauthor{A.~Acus \etal}

\institute{                    
  \inst{1} Vilnius University, Institute of Theoretical Physics and Astronomy, Go\v{s}tauto 12, Vilnius 01108, Lithuania\\
  \inst{2} Lithuanian University of Educational Sciences, Dept. Natural Science, Studentu 39, Vilnius 08106, Lithuania\\
  \inst{3} BLTP, JINR, Dubna, Russia\\
  \inst{4} Institute of Physics, University of Oldenburg, D-26111 Oldenburg, Germany\\
  \inst{5} Department of Theoretical Physics, Tomsk State Pedagogical University, Russia
}
\pacs{nn.mm.xx}{21.60.Fw}
\pacs{nn.mm.xx}{42.65.Hw}
\pacs{nn.mm.xx}{42.65.Wi}

\abstract{
We continue the discussion on the interaction energy of the axially
symmetric Hopfions evaluated directly from the product anzsatz.
The Hopfions are given by the projection of Skyrme model solutions onto
the coset space SU(2)/U(1). Our results show that if the
separation between the constituents is not small, the product ansatz can
be considered as a good approximation to the general
pattern of Hopfions interaction both in repulsive and attractive channel.}

\newcommand{\bphi}{\ensuremath{\boldsymbol{\phi}}}
\newcommand{\BU}{\ensuremath{\mathbf{U}}}
\newcommand{\BH}{\ensuremath{\mathbf{H}}}
\newcommand{\Tr}{\mathop{\mathrm{Tr}}}
\newcommand{\Br}{\ensuremath{\mathbf{r}}}
\newcommand{\BR}{\ensuremath{\mathbf{R}}}
\newcommand{\ii}{\mathrm{i}}

\begin{document}

\maketitle


\section{Introduction}

Topological and non-topological solitons appear in many non-linear field models in various contexts. Since the appearance of the soliton
solutions on the theoretical scene in the 1970s, it has become evident that they play a prominent role in
classical and quantum field theory. These spatially localized non-perturbative stable field configurations
are natural in a wide variety of physical systems \cite{Manton-Sutcliffe}.

The Faddeev-Skyrme model in $d=3+1$ is a
modified scalar $O(3)$-sigma model with a quartic in derivatives term \cite{Faddeev}. The structure
of the Lagrangian of this model is similar with the original Skyrme model \cite{Skyrme:1961vq}, whose solitons are
posited to model atomic nuclei, however the topological properties of the corresponding solitions, Hopfions and
Skyrmions, are very different. It was shown that soliton solutions of the Faddeev-Skyrme model
should be not just closed  flux-tubes of the fields  but {\it knotted\/} field configurations.

The first explicit non-trivial Hopfion solutions were
constructed numerically by Battye and Sutcliffe \cite{Battye1998} who found the trefoil knotted solution in the Faddeev-Skyrme
model. Consequent analyses revealed a very rich structure of the Hopfion states \cite{Hietarinta2000,Sutcliffe:2007ui}. The
subsequent development have revealed  a plethora of such  topological solutions with a non-trivial value of
the Hopf invariant, which play a prominent role in the modern physics \cite{Kaufmann}, chemistry \cite{MacArthur} and biology
\cite{Sumners}. A number of different models which describe topologically stable knots associated with the first Hopf map $S^3 \to
S^2$ are known in different contexts. It was argued, for example, that a system of two coupled  Bose condensates
may support Hopfion-like solutions \cite{Babaev}, or that
glueball configurations in QCD may be treated as Hopfions.

Note that most of the investigations of the Skyrmions and Hopfions mainly focus
on the search for classical static solutions.
Indeed, since in both models these configurations do not saturate the
topological bound, the powerful technique of the moduli space approximation cannot be  directly applied to analyse
the low-energy dynamics of the solitons. Therefore in order to investigate the process of interaction between these solutions
one has to implement rather advanced numerical methods.

Interestingly, the numerical simulations of the head-on collision
of the charge one Skyrmions still reveal the well known pattern of the
$\pi/2$ scattering through the intermediate axially-symmetric
charge two Skyrmion \cite{Battye:1996nt}, which is typical for self-dual
configurations like BPS monopoles \cite{Manton-Sutcliffe}. However recent attempt to model the
Hopfion dynamics \cite{Hietarinta:2011qk} failed to find the
channel of right-angle scattering in head-on collisions of the charge one solitons.

Another approach to the problem of interaction between the stringlike solitons of the Faddeev-Skyrme model
is to consider the asymptotic fields of the Hopfion of degree one, which corresponds
to a doublet of orthogonal dipoles \cite{Gladikowski:1996mb,Ward:2000qj}. Investigating this limit Ward
predicted existence of three attractive channels in the interaction of the
Hopfions with different orientation \cite{Ward:2000qj}.

In his pioneering paper \cite{Skyrme:1961vq} Skyrme suggested to implement
so-called product ansatz to approximate a composite configuration of well-separated individual Skyrmions.
The ansatz is constructed by the multiplication of
the Skyrmion matrix-valued fields. Note that besides the rational map ansatz \cite{Houghton:1997kg} it can be applied to
produce an initial multi-Skyrmion configuration for consequent numerical calculations in a sector of given degree \cite{Battye1998}.
Evidently, the same approach can be used to model the configuration of well separated static Hopfions of degree one
to approximate various multicomponent configurations.

Recently we discussed
the relation between the solutions of the Skyrme model of lower degree and the corresponding axially
symmetric Hopfions which is given by the projection onto the coset space $SU(2)/U(1)$ \cite{Acus:2014tqa}.
Using this approach we made use of the product ansatz of two well-separated single Hopfions and
confirmed that the product ansatz correctly reproduces the channels of interaction between them. In this paper
we briefly describe the relation between the solutions of the Skyrme model
of lower degrees and the corresponding axially symmetric Hopfions which is given by the projection onto the coset space,
adding here new and updated results of the numerical evaluation of the corresponding interaction energy of the Hopfions.

\section{The model}
The Faddeev-Skyrme model in 3+1 dimensions with metric $(+,-,-,-)$ is defined by the Lagrangian
\begin{equation}
\label{model}
{\cal L} = \frac{1}{32\pi^2}\left(\partial_\mu \phi^a \partial^\mu \phi^a -
\frac{1}{4}(\varepsilon_{abc}\phi^a\partial_\mu \phi^b\partial_\nu \phi^c)^2 \right)\,,
\end{equation}
where $\phi^a = (\phi^1, \phi^2,\phi^3)$ denotes a triplet of scalar real
fields which satisfy the constraint $|\phi^a|^2=1$.
The finite energy configurations approach a constant value at spatial infinity, which
we choose to be $\phi^a(\infty) = (0,0,1)$. For fields $\bphi(\mathbf{x})$ with this property the domain of definition, i.~e. 3D Euclidian space, is equivalent to $S^3$ sphere and $\bphi(\mathbf{x})$ defines the map $S^3 \to S^2$. It is well known that these maps are characterized by Hopf invariants $Q = \pi_3(S^2) = \mathbb{Z}$, where the
target space $S^2$ by construction is the coset space $SU(2)/U(1)$.

Any coset space element $\BH$ can be projected
from generic $SU(2)$ group element $\BU$. In circular coordinate system the projection takes the form
\begin{equation}
\label{genericProjection}
\BH=2\sum_a (-1)^a \tau_a \phi_{-a}= 2 \BU \tau_0 \BU^\dagger\,,
\end{equation}
where the Pauli matrices $(\tau_1,\tau_0,\tau_{-1})$ satisfy relation
\begin{equation}
\label{pauliDefinition}
\tau_a \tau_b =\frac14 (-1)^a \delta_{a,-b}\mathbf{1} -\frac{1}{\sqrt{2}}
\left[
\begin{array}{ccc}
1 & 1 &1\\
a & b & c
\end{array}
\right]\tau_c .
\end{equation}
The symbol in square brackets denotes the Clebsch-Gordon coefficient. Then the Lagrangian~(\ref{model})
can be rewritten in terms of coset space elements $\BH$,
\begin{equation}
\label{modelInH}
\begin{split}
{\cal L} = &\frac{1}{64\pi^2}\Bigl(\Tr\big\{\partial_\mu \BH \partial^\mu \BH\big\} \\&
+
\frac{1}{16}\Tr\big\{\bigl[\partial_\mu \BH,\partial_\nu \BH\bigr]\bigl[\partial^\mu \BH,\partial^\nu \BH\bigr]\big\} \Bigr)\,.
\end{split}
\end{equation}

The difference between the Skyrmions and Hopfions is that in the latter case the dimensions of the domain space
and the target space are not the same. The topological charge of the Hopfions, which meaning is the
linking number in the domain space \cite{Faddeev}, is not defined locally.

There have been many investigations of the solutions of the model~(\ref{model}) for higher degree $Q$
\cite{Gladikowski:1996mb,Sutcliffe:2007ui,Battye1998,Hietarinta2000}. Here we consider projections of
general rational map Skyrmion ansatz, however numerical results of interaction potential is
presented only for axially symmetric configurations of lower degrees $Q=1,2$ which are conventionally labeled as
${\cal A}_{1,1}$ and ${\cal A}_{2,1}$ \cite{Sutcliffe:2007ui}.

An rational map approximation to these solutions can be constructed via Hopf
projection of the corresponding Skyrmion configurations~\cite{Su:2008}. Recall that the rational map ansatz~\cite{Houghton:1998kg} is an approximation
to the ground state solution of the Skyrme model, which for baryon number $B\ge 1$ takes the following form:
\begin{equation}\label{RatMapGen}
U_{R}(\mathbf{r})=\exp (2\mathrm{i}\, \hat{n}^{a}_R\hat{\tau}_{a}F(r))\, .
\end{equation}
The unit vector $\hat{\mathbf{n}}_R$ is defined
in terms of a rational complex function $R(z)=p(z)/q(z)$, where  $p(z)$ and $q(z)$
are polynomials of complex variable $z$ of degree at most $N$, and $p$ and $q$ have no common roots. In Cartesian coordinates
the components of $\hat{{\mathbf n}}_R$ then can be written as
\begin{equation}\label{nInSpher}
\hat{{\mathbf n}}_R=\frac{1}{1+|R|^2} \{2 \Re(R), 2 \Im(R), 1-|R|^2\}\, .
\end{equation}
Parametrizing the complex variable $z=\tan (\theta/2) \mathrm{e}^{\mathrm{i}\varphi}$ by polar and
azimuthal angles $\theta$ and $\varphi$ we can find explicit expression for any given rational function $R$.
In particular, for baryon charge $B=1$ we take $R(z)=z$, which gives
\begin{equation}\label{Reqz}
\hat{{\mathbf n}}_{1}=\hat{{\mathbf r}}=\{\bigl(\mathrm{e}^{-\mathrm{i}\varphi}\sin\theta\bigr)/\sqrt{2},\cos\theta ,
\bigl(-\mathrm{e}^{\mathrm{i}\varphi}\sin\theta\bigr)/\sqrt{2}\}\, .
\end{equation}
We also implement the circular coordinates  $\hat{n}_{\pm 1}=\mp\frac{1}{\sqrt{2}}\bigl(\hat{n}^1\pm\mathrm{i}\hat{n}^2\bigr)$
and $\hat{n}_0=\hat{n}^3$, where $\hat{n}^i$ denotes the Cartesian components of the unit vector (\ref{nInSpher}).
It is known that simple choice $R(z)=z$  yields an exact solution of the model.

For $B=2$ Skyrmion the lowest energy rational map approximation is given by the choice $R(z)=z^2$, or explicitly
in circular coordinates
\begin{equation}
\hat{{\mathbf n}}_{2}=\{\frac{\mathrm{e}^{-2\mathrm{i}\varphi}\sin^2\theta}{\sqrt{2}\bigl(1+\cos^2\theta\bigr)}
,
\frac{2\cos\theta}{1+\cos^2\theta}
,
\frac{-\mathrm{e}^{2\mathrm{i}\varphi}\sin^2\theta}{\sqrt{2}\bigl(1+\cos^2\theta\bigr)}
\}\, .
\end{equation}
For higher baryon numbers rational map approximations are also well known~\cite{Manton-Sutcliffe}, thought the
correspondence between them and exact numerical results is getting
worse as baryon number $B$ increases. Note, that in the topologically trivial sector $B=0$  we may take
$\hat{{\mathbf n}}_{0}=\{1/\sqrt{2},0, -1/\sqrt{2}\}$, which we used to test the product ansatz approximation.

Note that the Skyrme model can be consistently reduced~\cite{Su:2008} to Faddeev-Skyrme model by restricting SU(2) Lie
algebra currents $\BU^\dagger\partial_\mu\BU$ to coset representation~(\ref{genericProjection}). This means that
Faddeev-Skyrme fields configurations with Hopf charge $Q$ corresponds to Skyrme field with the baryon number $B$.
Therefore the projection of rational map ansatz~(\ref{RatMapGen}) yields the rational map approximation of the Hopf charge of
the same degree  $Q=N$. Moreover, the usual profile function $F(r)$ of Skyrme model, which is a monotonically decreasing
function satisfying boundary conditions $F(0)=\pi,~~F(\infty)=0$, can be used. This projection produces the Hopfion configuration
of degree one with mass $1.232$.

As usual, we denote the ${\cal A}_{1,1}$ configuration as $\BH_1$ which is a projection
of the Skyrmion matrix valued field $\BU_{R=z}$, i.e.
\begin{equation}
\label{hopfion1Projection}
\BH_1(\Br)=2\BU_{R=z}(\Br) \tau_0 \BU_{R=z}^\dagger(\Br)\,,
\end{equation}
where $\BU_{R=z}(\Br)$ is the usual spherically symmetric hedgehog ansatz parametrised by rational
map~(\ref{RatMapGen}) with $R(z)=z$ and $\hat{\mathbf{n}}_R$ is defined by eq. (\ref{Reqz}). It should be noted, that
although the ansatz~(\ref{RatMapGen}) for $B=1$ is spherically symmetric, the corresponding Hopfion of degree
$Q=1$ does not possess the spherical symmetry. The projection breaks it down to axial symmetry ${\cal A}_{1,1}$ \cite{Battye1998}.

The position curve of Hopfion is chosen to be the curve of the preimages of the point $(0,0,-1)$ which is the
antipodal to the vacuum $(0,0,1)$. For the simplest ${\cal A}_{1,1}$ Hopfion this is a circle of radius
$F(r_c) = \pi/2$, with numerical value $r_c = 0.8763$ in the $x-y$ plane. Small deviations $F(r) = F(r_0)+\epsilon$ then define the
tube around the position curve where $\vartheta \approx \pi/2$ and $\varphi = [0,2\pi)$. The same is true for $R(z)=z^2$
or ${\cal A}_{2,1}$ configuration, except that point on tube rotates twice when angle $\varphi$ changes from $0$ to $2\pi$ and
the radius of circle being slightly large, $r_c = 1.299$.

For single Hopfion we can also rotate the points on the tube about the vertical $x_3$ axis by applying rotation transform via the $SU(2)$ matrix
\begin{equation}
\label{hopfionRotationM}
\BH\rightarrow  D(\alpha)\BH D(-\alpha)\, .
\end{equation}
This global transformation corresponds to the symmetry of the Lagrangian (\ref{modelInH}).

\begin{figure}[hbt]
\label{fig:1}
\begin{center}
\includegraphics[height=.27\textheight, angle =0]{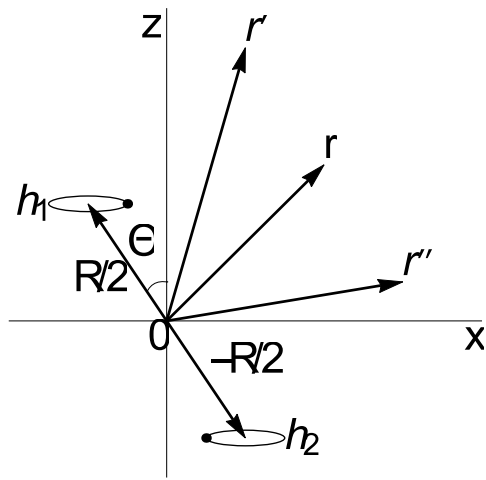}
\end{center}
\caption{\small Geometry of the system of two interacting Hopfions
$h_1$ and $h_2$, which are located at the points $\BR/2$ and $-\BR/2$ (indicated by arrows), respectively. The dot position on hopfion core illustrates the hopfion orientation (oposite phase case $\Delta \alpha =\pi$ is shown here).}
\end{figure}

Let us now consider two Hopfions of arbitrary charges which are placed
at the points $\BR/2$ and  $-\BR/2$ and separated by a distance $R$, as shown in Figure~\ref{fig:1}.
There the polar angle $\Theta$ corresponds to the orientation of the Hopfions with respect to the $z$-axis.
Note that for $R=z$ the pattern of interaction between the charge one Hopfions is invariant with respect to
the spacial rotations of the system around the $z$-axis by an azimuthal angle $\Phi$. This additional symmetry does
not appear for higher maps, e.g. for $R=z^2$.

First, we suppose that both separated Hopfions are of positive charges and they are in phase, i.e.  for example,
rotation matrices $D(\alpha)$ in (\ref{hopfionRotationM}) are identities for both Hopfions. More strictly,
the definition "in phase" only means that the difference between the angles of rotations of individual
Hopfions is zero, $\Delta \alpha =0$.

Then the system of two hopfions can be approximated by the product ansatz
\begin{equation}
\label{hopfionProductAnsatz}
\BH_N^{\Delta \alpha =0}(\Br)=2 \BU_{R^\prime}(\Br^\prime)\BU_{R^{\prime\prime}}(\Br^{\prime\prime})\tau_0\BU_{R^{\prime\prime}}^\dagger(\Br^{\prime\prime})\BU_{R^\prime}^\dagger(\Br^\prime)\,,
\end{equation}
where
$\Br^\prime=\Br+\BR/2$ and $\Br^{\prime\prime}=\Br-\BR/2$ and $N$ denotes the sum of degrees of rational maps $R^\prime$ and ${R^{\prime\prime}}$.
Fields of both Hopfions in (\ref{hopfionProductAnsatz})
at the spacial boundary tend to the same asymptotics $(0,0,1)$. Also note, that in the constituent
system (\ref{hopfionProductAnsatz}) of two Hopfions, contrary to the single Hopfion case, the transformation (\ref{hopfionRotationM}) of one
of the Hopfions $\BH$ do not leave the Lagrangian (\ref{modelInH}) invariant, it becomes a function of the relative
phase $\Delta\alpha$.

In addition to the ansatz (\ref{hopfionProductAnsatz}) we can as well consider the system of two separated Hopfions, when one
of them is relatively rotated by arbitrary angle $\alpha$. Here, hovewer, we only restrict ourself to the relative phase
$\Delta \alpha =\pi$, i.e., to the case when Hopfions have opposite phases. Taking one of rotation matrix $D(0)$ to be
identity matrix and the other $D(\pi)=2\tau_0$,
we can express this system in terms of the matrices $\BU_{R^\prime}$ and $\BU_{R^{\prime\prime}}$, thus the corresponding product ansatz is
now different from (\ref{hopfionProductAnsatz}):
\begin{equation}
\label{hopfionProductAnsatzAntiParallel}
\BH_N^{\Delta \alpha =\pi}(\Br)=8
\BU_{R^\prime}(\Br^\prime)\tau_0\BU_{R^{\prime\prime}}(\Br^{\prime\prime})\tau_0\BU_{R^{\prime\prime}}^\dagger(\Br^{\prime\prime})\tau_0\BU_{R^\prime}^\dagger(\Br^\prime)\,,
\end{equation}

The product ansatz approximations (\ref{hopfionProductAnsatz}) and (\ref{hopfionProductAnsatzAntiParallel}) ensures the conservation of
the total topological charge for any separation $R$ and space orientation of the constituents. Note, however, that in the case
of different rational maps ${R^\prime}\neq R^{\prime\prime}$, for example, $z$ and $z^2$, which are considered below,  their order
in (\ref{hopfionProductAnsatz}) and (\ref{hopfionProductAnsatzAntiParallel}) is  important. The different ordering yields different
numerical results for small separation distances $R$, because the chiral angles for
$F_{R^\prime}(r^\prime)$ and $F_{R^{\prime\prime}}(r^{\prime\prime})$ differ. This problem will be investigated in more detail elsewhere.

Substitution of product ansatzes (\ref{hopfionProductAnsatz}) and (\ref{hopfionProductAnsatzAntiParallel})
into Lagrangian (\ref{modelInH}) yields energy densities
of both  configurations in terms of components of the position vectors
$r_i^{\prime}$ and $r_j^{\prime\prime}$ (cf Figure~\ref{fig:1}).

Let us express these components via Hopfion's position coordinates
$R,\Theta,\Phi$ and the spherical coordinates $r,\theta,\varphi$, then the numerical integration of the corresponding
local densities  over the variables $\varphi,\vartheta$ and $r$ yields the total energy (mass) of the system and
its topological charge. The density of the latter quantity in circular coordinates is given by the trace formula
\begin{equation}
\label{barDensGen}
\begin{split}
{\cal Q}(\Br^{\prime},\Br^{\prime\prime})&=\ii \sqrt{2}(-1)^{a+b} \left[
\begin{array}{ccc}
1 & 1 &1\\
a & b & a+b
\end{array}
\right]\\
&
\times\mathop{\mathrm{Tr}}\Bigl(
\nabla_a \bigl(\BU_{R^\prime}(\Br^{\prime})\BU_{R^{\prime\prime}}(\Br^{\prime\prime})\bigr) \BU_{R^{\prime\prime}}^\dagger(\Br^{\prime\prime})\BU_{R^\prime}^\dagger(\Br^{\prime})\nonumber\\
&\times\nabla_b \bigl(\BU_{R^\prime}(\Br^{\prime})\BU_{R^{\prime\prime}}(\Br^{\prime\prime})\bigr) \BU_{R^{\prime\prime}}^\dagger(\Br^{\prime\prime})\BU_{R^\prime}^\dagger(\Br^{\prime})
\\
&\times\nabla_{-a-b} \bigl(\BU_{R^\prime}(\Br^{\prime})\BU_{R^{\prime\prime}}(\Br^{\prime\prime})\bigr) \BU_{R^{\prime\prime}}^\dagger(\Br^{\prime\prime})\BU_{R^\prime}^\dagger(\Br^{\prime})
\Bigr)\,.\nonumber
\end{split}
\end{equation}

We used the evaluation of the total topological charge of the configuration (for different values of the separation
parameter $R$) as a correctness test
of our numerical computations.
The potential energy of interaction between two Hopfions of degree one and two is evaluated by subtracting of the
corresponding masses of single Hopfions, i.e.
$m_{(R^\prime=z)}=1.2314$ and  $m_{(R^{\prime\prime}=z^2)}= 2.079503$, from the integrated
density (\ref{modelInH}) evaluated on the configurations (\ref{hopfionProductAnsatz}) and (\ref{hopfionProductAnsatzAntiParallel}),
respectively.

\section{Numerical results}
\begin{figure}[hbt]
\begin{center}
a)\includegraphics[height=.20\textheight,angle =0]{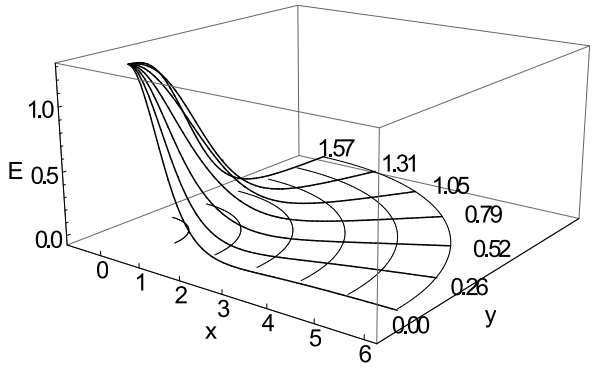}\newline
\end{center}
\vspace{-0.5cm}
\caption{\small The evaluated interaction energy of the $\Delta \alpha =0$ (in phase), product $Q=1+1$ ansatz Hopfions as a function of the orientation parameters $R$ and $\Theta$.}\label{fig:2}
\end{figure}

Evaluation of the total topological charge and the energy of the product ansatz configuration  requires numerical integration.
In particularly, for each given set of fixed values
of the orientation parameters $R$, $\Theta$ and $\Phi$, the integration of the energy density
and Hopfion charge density (\ref{barDensGen}) over three components of the Hopfion field  yields, correspondingly,
the strength of the interaction energy of the Hopfions and the topological charge of the configuration.

\begin{figure}[hbt]
\begin{center}
\hspace{0.5cm} a)\hspace{-0.6cm}
\includegraphics[height=.23\textheight, angle =0]{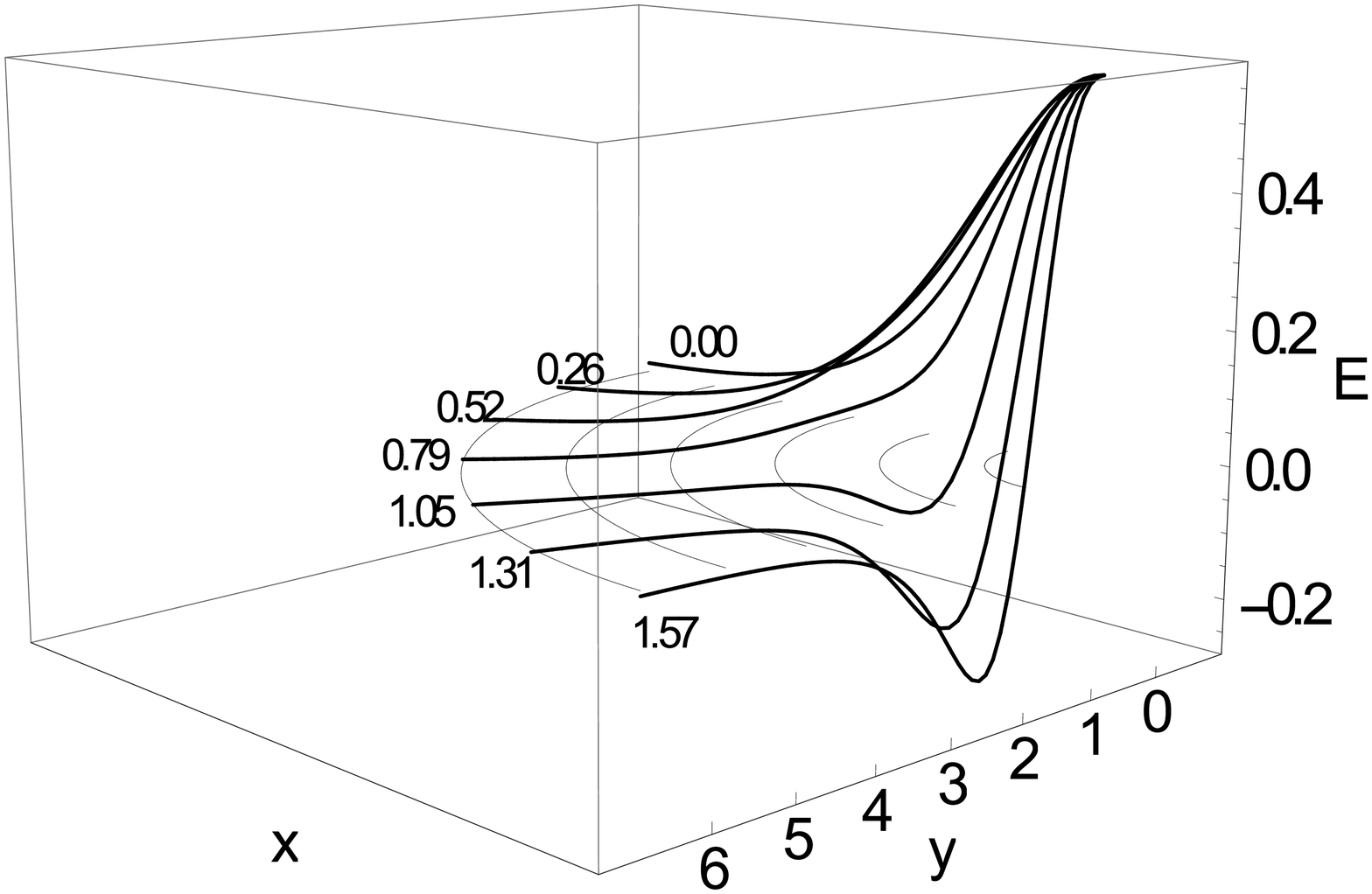}
\end{center}
\vspace{-0.5cm}
\caption{\small The interaction energy of the $\Delta \alpha =\pi$ (opposite phases), product $Q=1+1$ ansatz Hopfions  as a function of the orientation parameters $R$ and $\Theta$.}\label{fig:3}
\end{figure}

\begin{figure}[hbt]
\begin{center}
a)\hspace{-0.6cm}
\includegraphics[height=.22\textheight,angle =0]{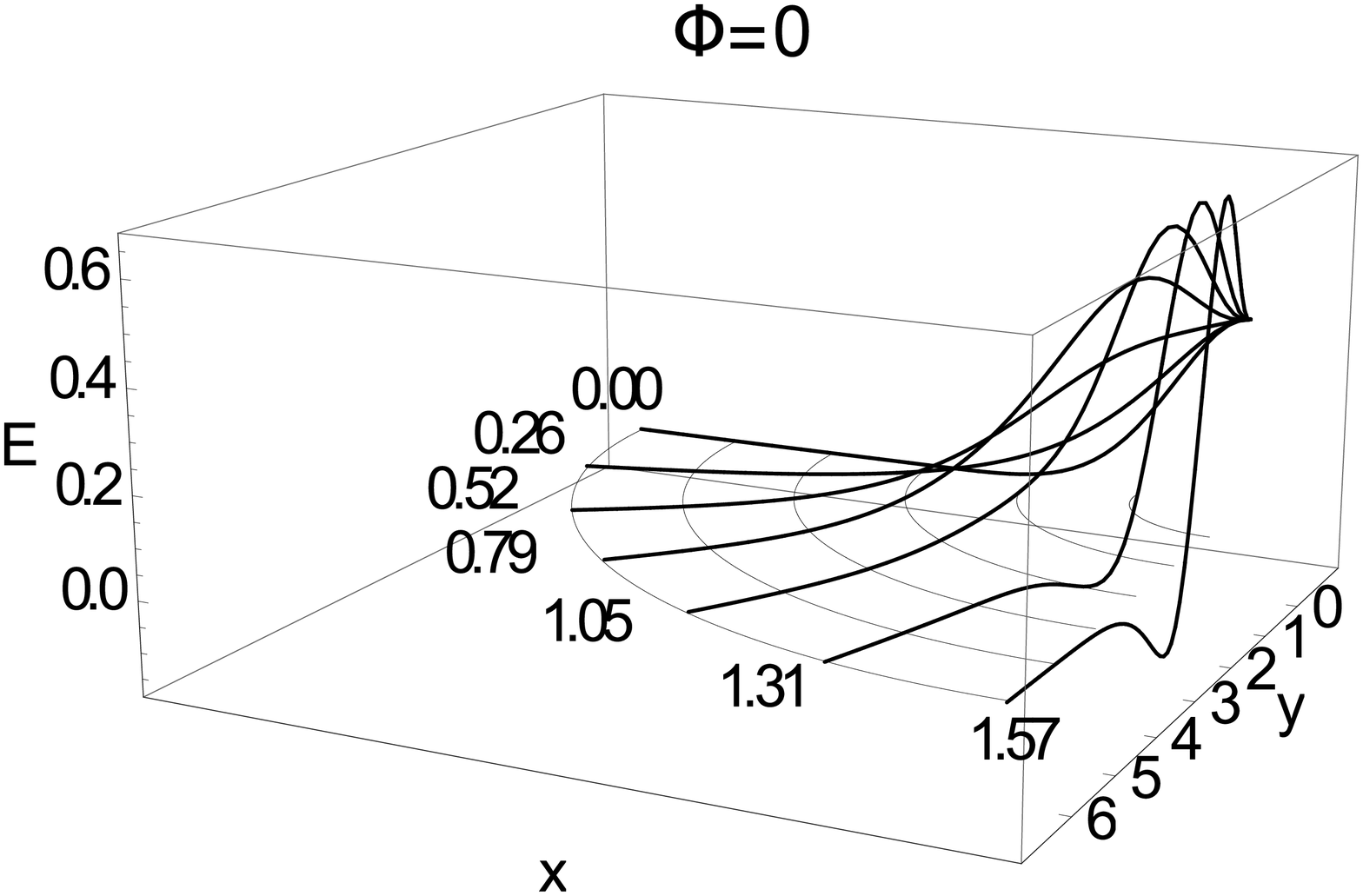}
b)\hspace{-0.6cm}
\includegraphics[height=.22\textheight, angle =0]{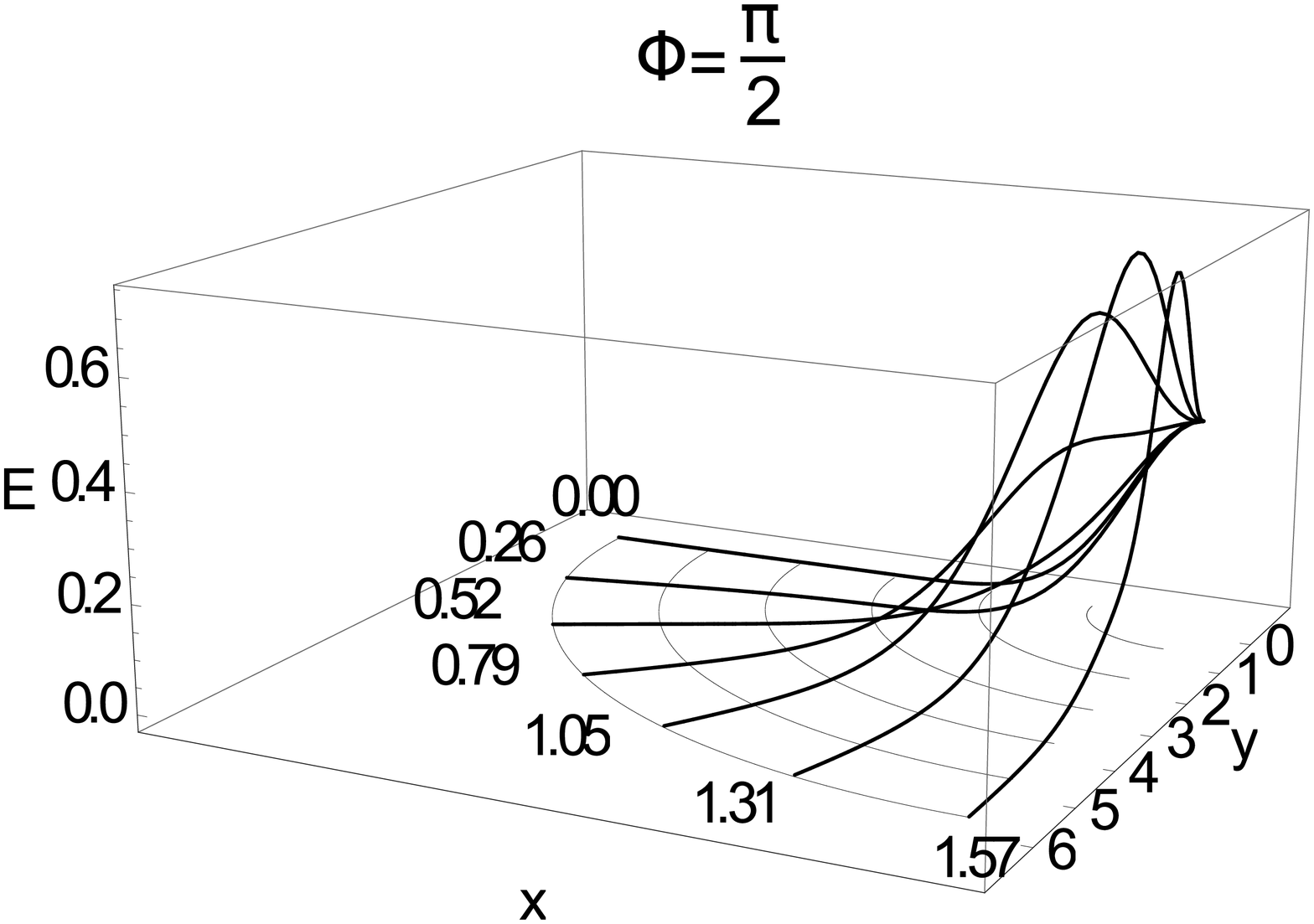}

\end{center}
\vspace{-0.5cm}
\caption{\small The evaluated interaction energy of the $\Delta \alpha =0$ (in phase),
 $Q=1+2$ product ansatz Hopfions is plotted as function of the orientation parameters $R$ and $\Theta$
 for fixed angles $\Phi=0$ a) and $\Phi=\pi/2$ b).
}\label{fig:4}
\end{figure}

\begin{figure}[hbt]
\begin{center}
a)\hspace{-0.6cm}
\includegraphics[height=.21\textheight,angle =0]{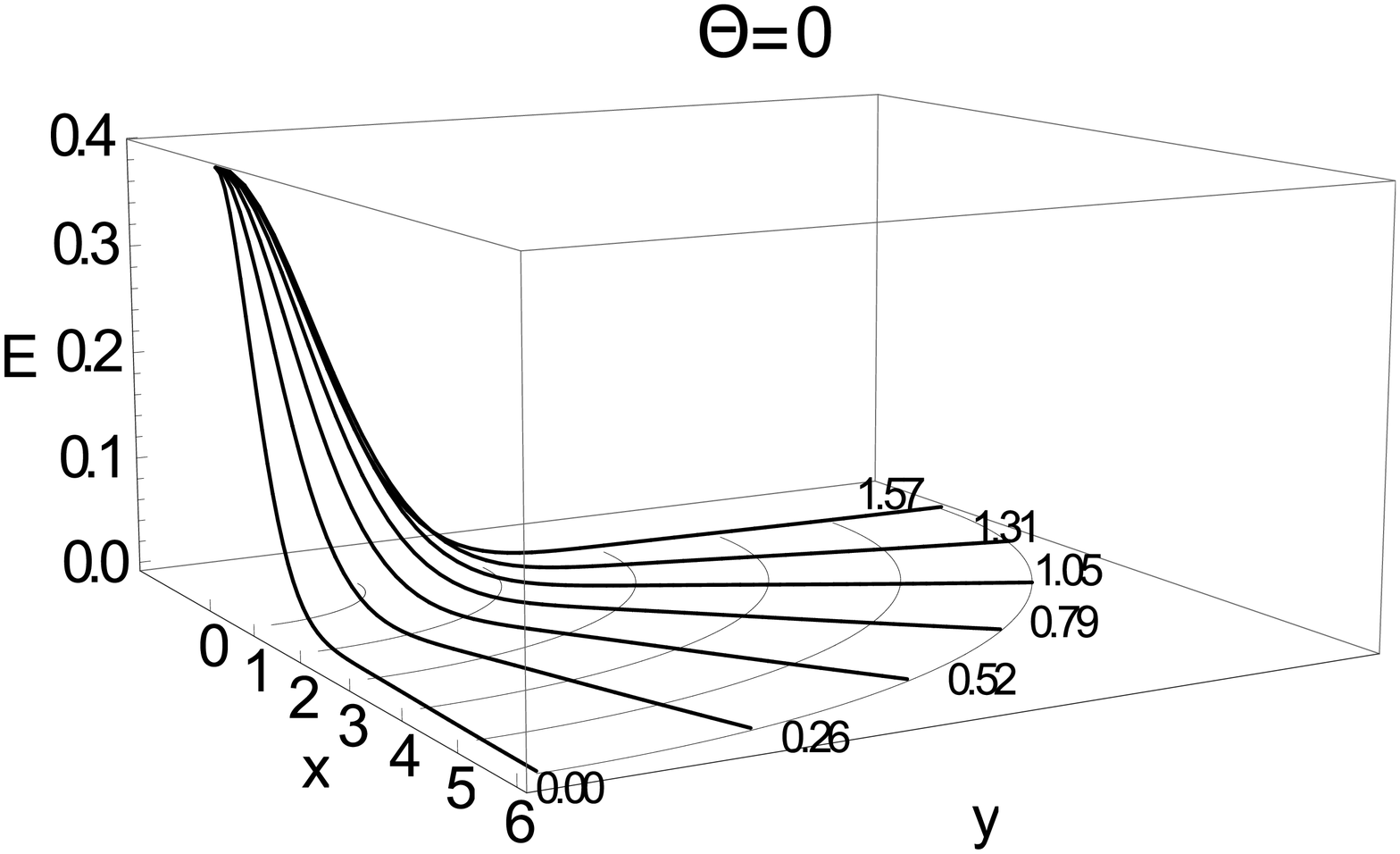}
b)\hspace{-0.6cm}
\includegraphics[height=.21\textheight, angle =0]{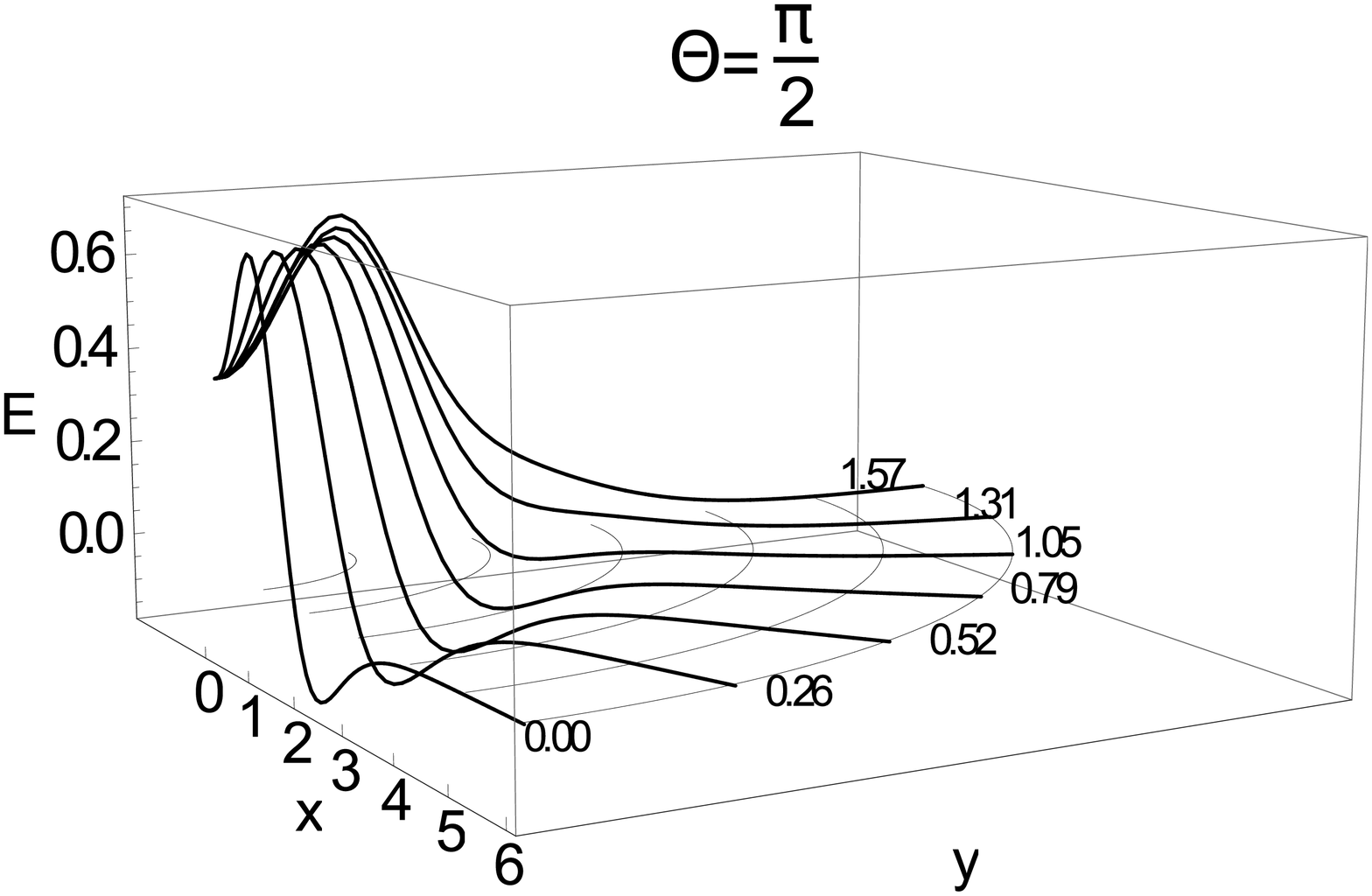}
\end{center}
\vspace{-0.5cm}
\caption{\small The evaluated interaction energy of the $\Delta \alpha =0$
(in phase),  $Q=1+2$ product ansatz Hopfions as a function of the orientation parameters $R$ and
$\Phi$ for fixed angles $\Theta=0$ a) and $\Theta=\pi/2$ b).
}\label{fig:5}
\end{figure}

\begin{figure}[hbt]
\begin{center}
a)\hspace{-0.6cm}
\includegraphics[height=.21\textheight,angle =0]{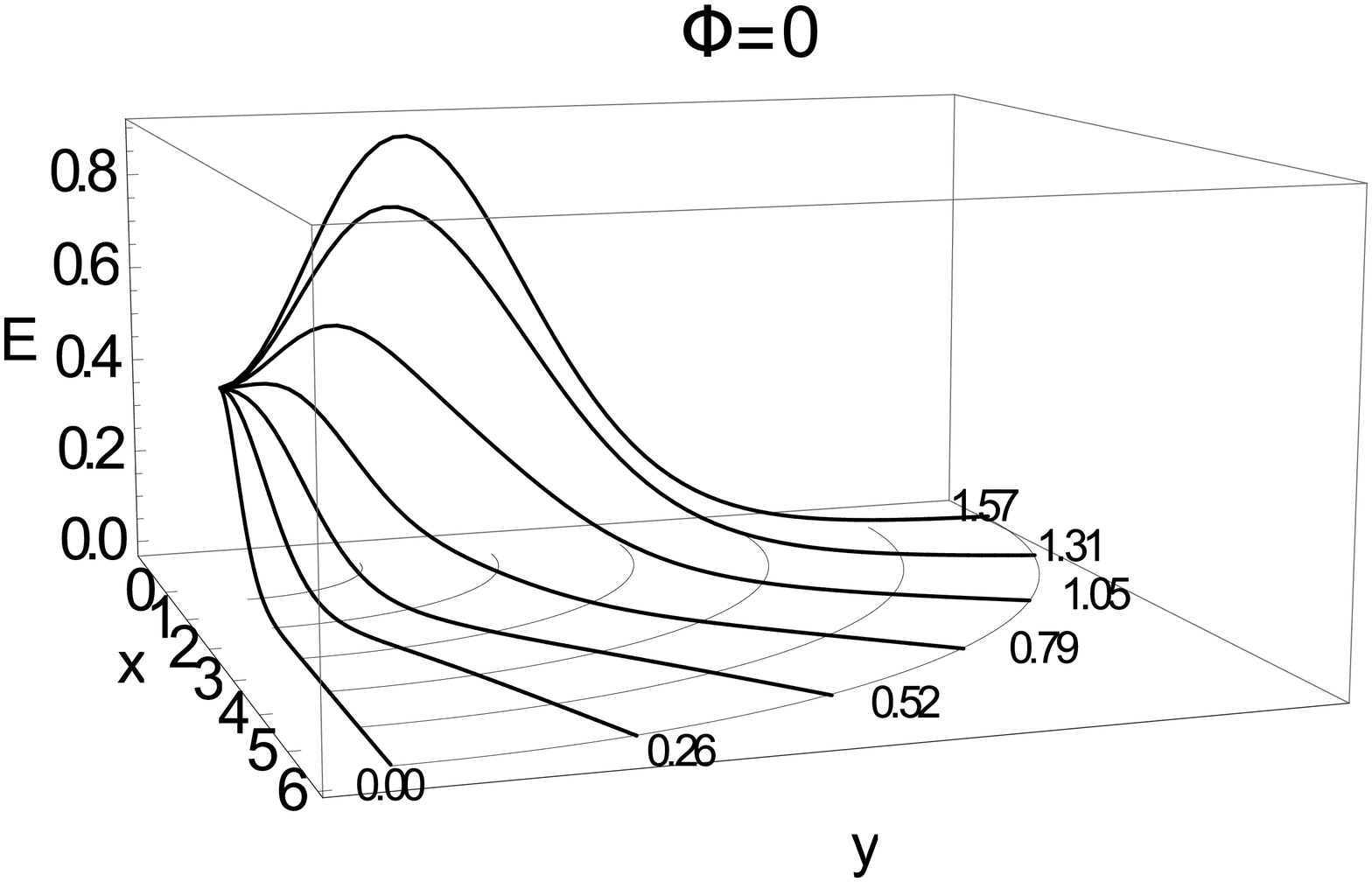}
b)\hspace{-0.6cm}
\includegraphics[height=.21\textheight, angle =0]{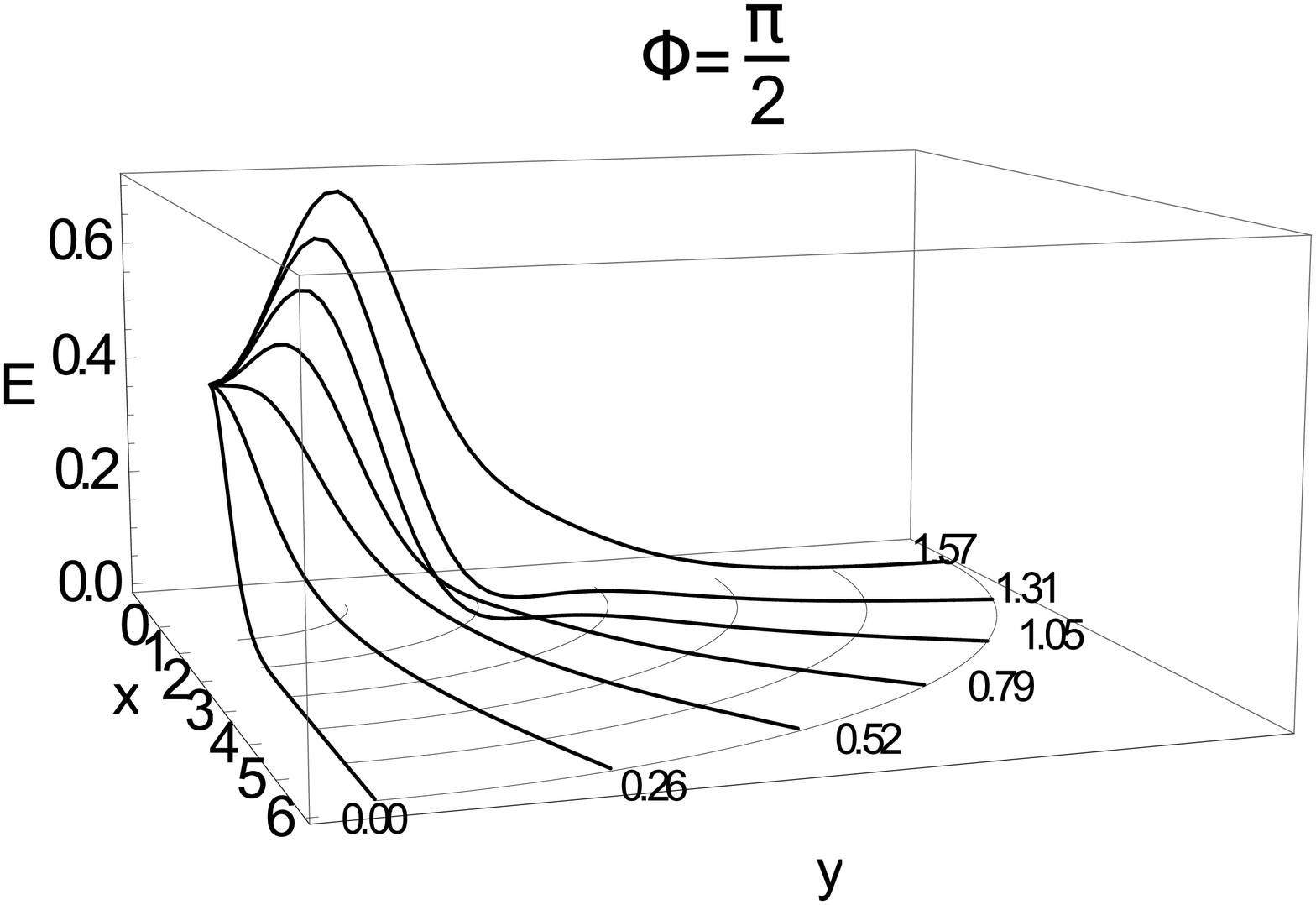}
\end{center}
\vspace{-0.5cm}
\caption{\small The evaluated interaction energy of the $\Delta \alpha =\pi$ (opposite phase),  $Q=1+2$ product ansatz Hopfions
as a function of the orientation parameters $R$ and $\Theta$ for fixed angles $\Phi=0$ a) and $\Phi=\pi/2$ b).
}\label{fig:6}
\end{figure}

\begin{figure}[hbt]
\begin{center}
a)\hspace{-0.6cm}
\includegraphics[height=.2\textheight,angle =0]{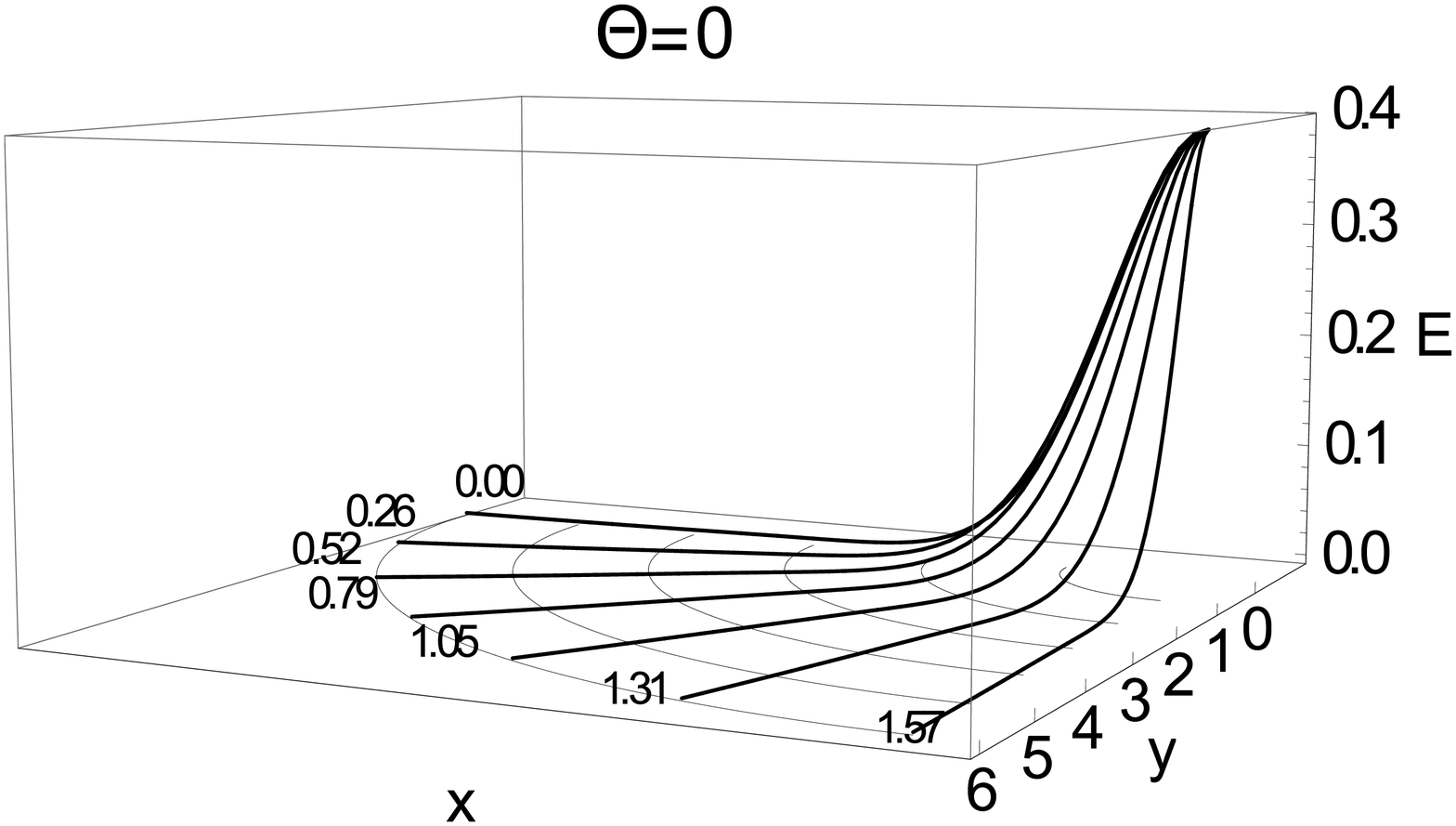}
b)\hspace{-0.6cm}
\includegraphics[height=.2\textheight, angle =0]{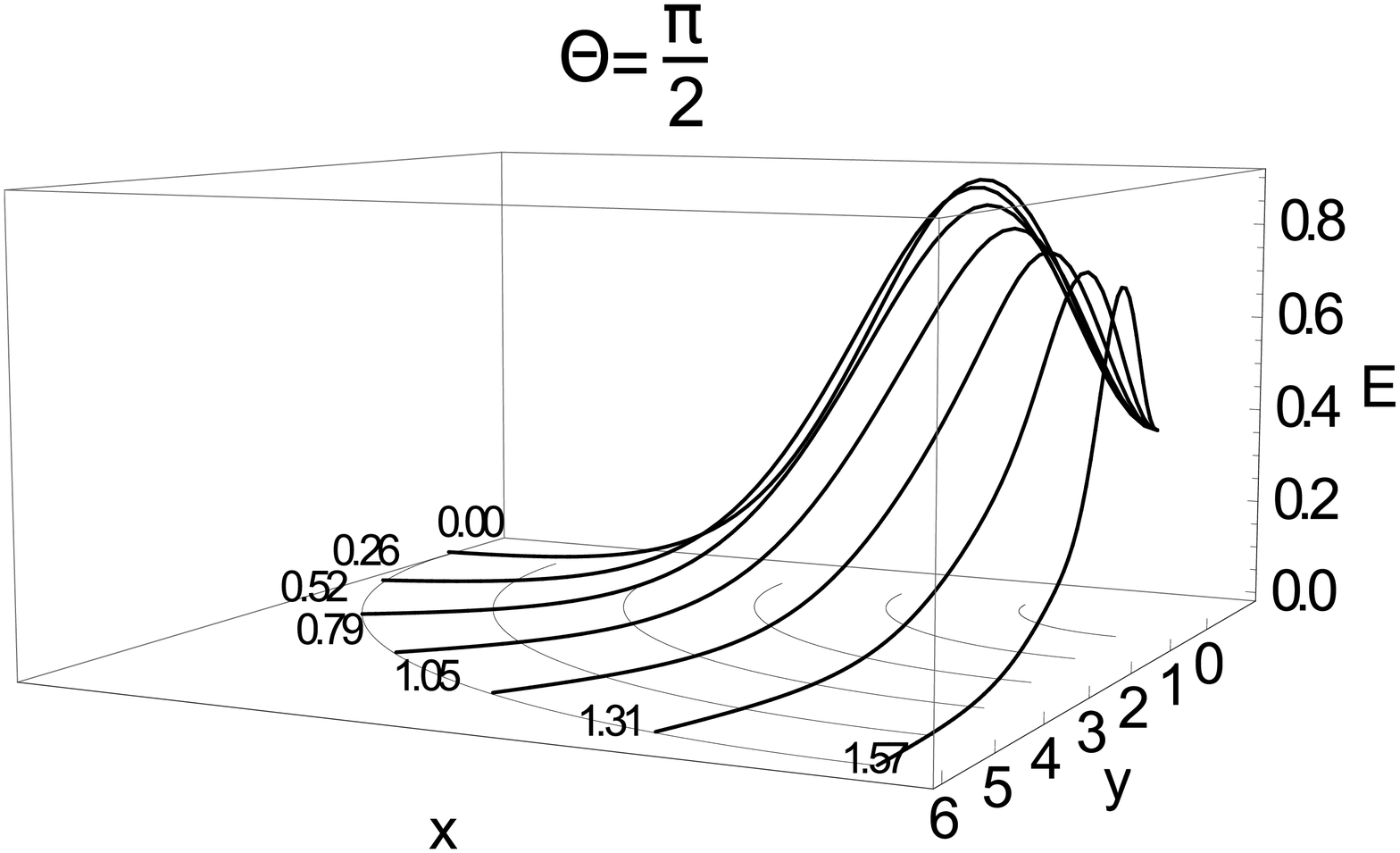}
\end{center}
\vspace{-0.5cm}
\caption{\small The evaluated interaction energy of the $\Delta \alpha =\pi$ (opposite phase),  $Q=1+2$ product ansatz Hopfions as a
function of the orientation parameters $R$ and $\Phi$ for fixed angles $\Theta=0$ a) and $\Theta=\pi/2$ b).
}\label{fig:7}
\end{figure}

\begin{figure}[hbt]
\begin{center}
a)\hspace{-0.6cm}
\includegraphics[height=.21\textheight,angle =0]{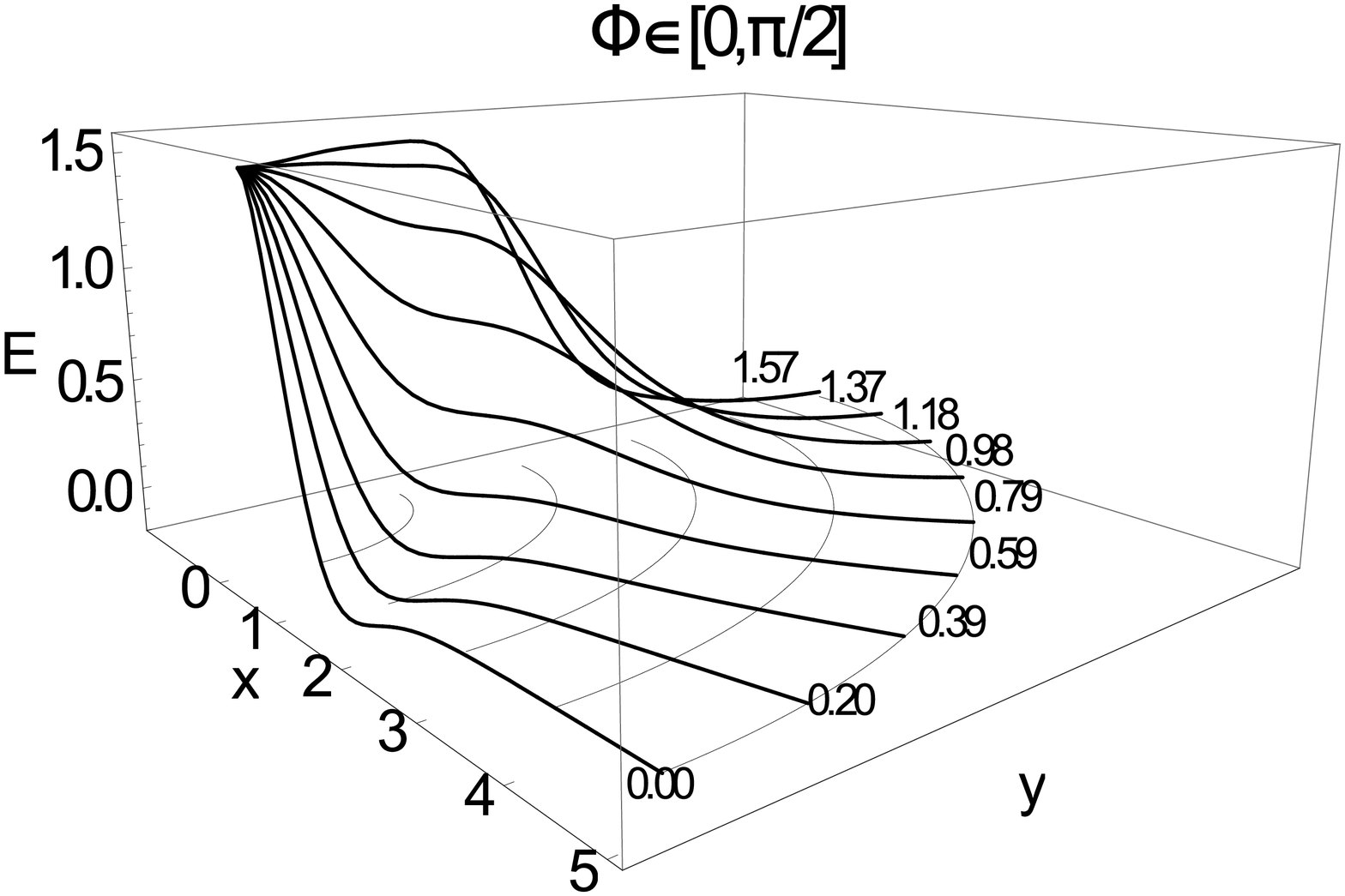}
b)\hspace{-0.6cm}
\includegraphics[height=.21\textheight, angle =0]{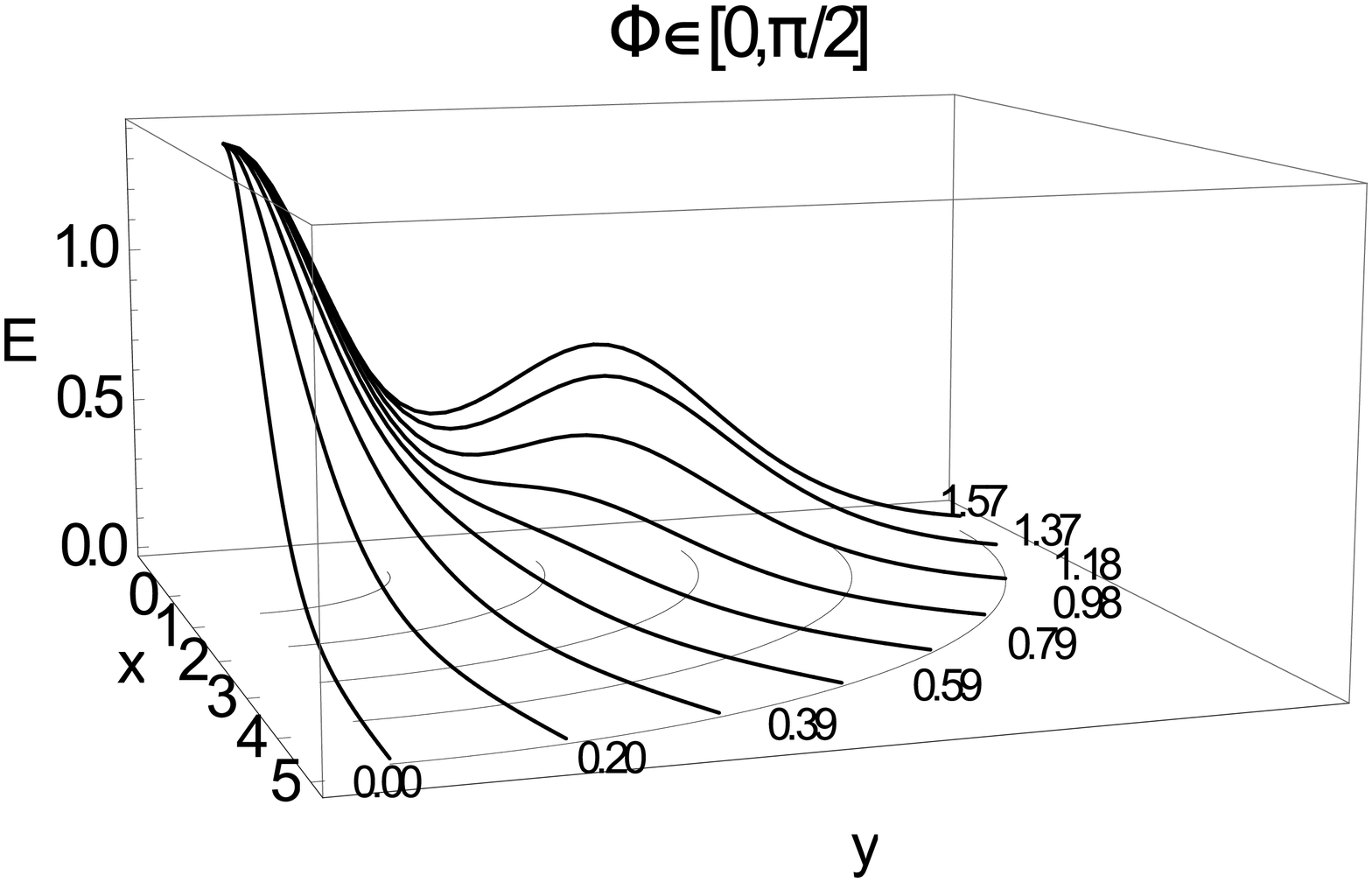}
\end{center}
\vspace{-0.5cm}
\caption{\small The evaluated interaction energy of the $\Delta \alpha =0$ (in phase)~a) and of the $\Delta \alpha =\pi$ (opposite phase)~b) of
 $Q=2+2$ product ansatz Hopfions as a function of the orientation parameters $R$ and $\Theta$ for angles $\Phi\in [0,\pi/2]$.
}\label{fig:8}
\end{figure}

\begin{figure}[hbt]
\begin{center}
a)\hspace{-0.6cm}
\includegraphics[height=.2\textheight,angle =0]{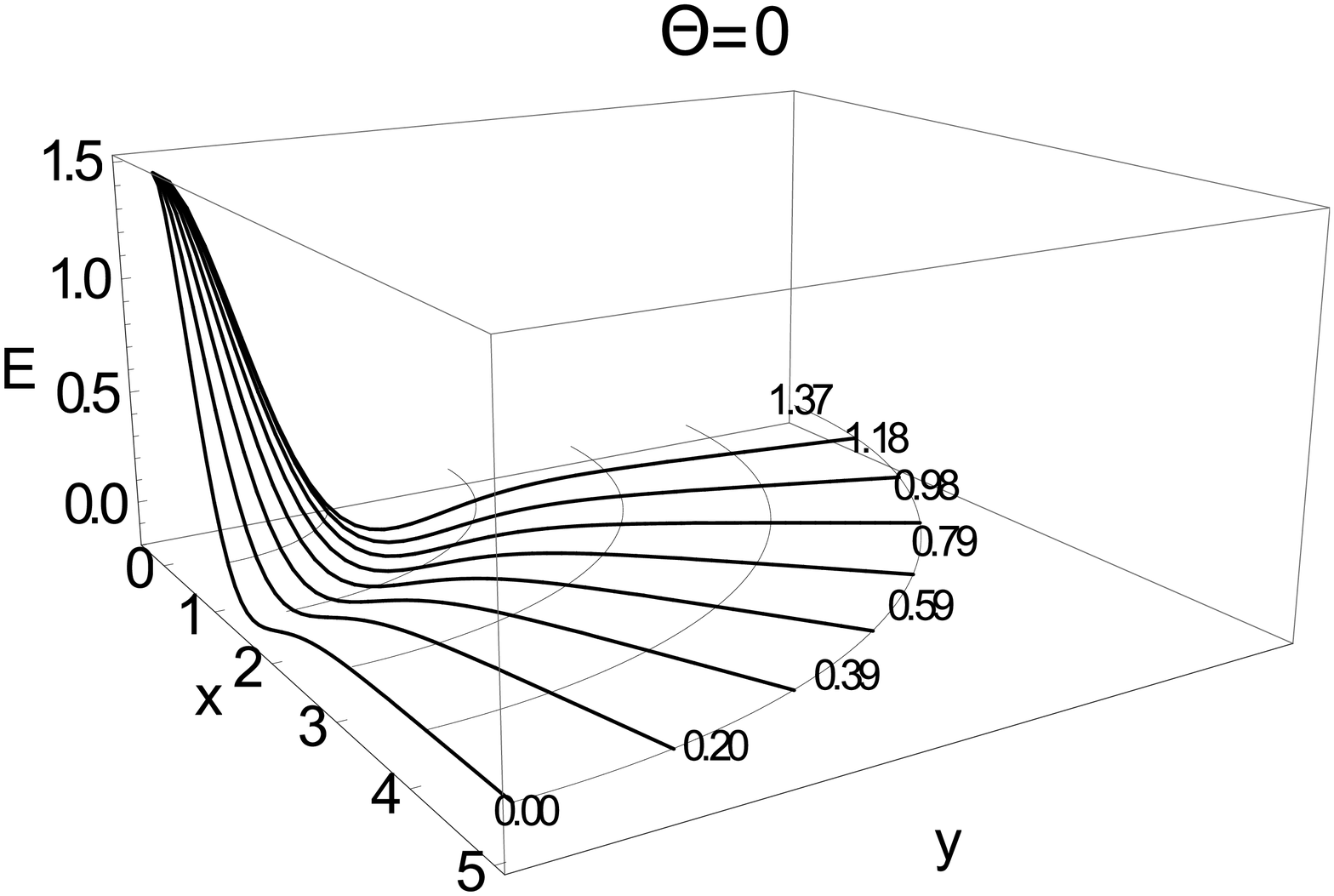}
b)\hspace{-0.6cm}
\includegraphics[height=.2\textheight, angle =0]{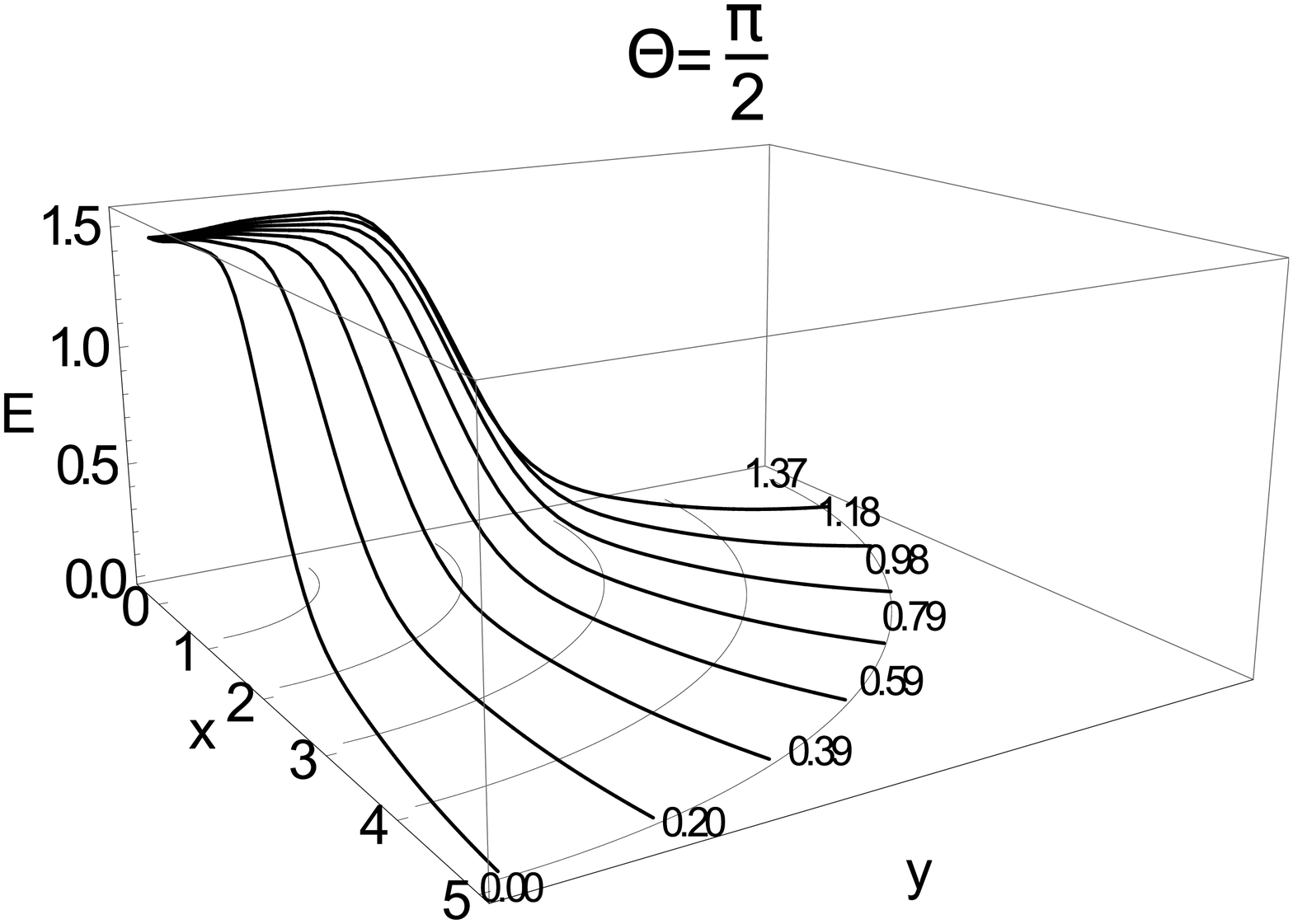}
\end{center}
\vspace{-0.5cm}
\caption{\small The evaluated interaction energy of the $\Delta \alpha =0$ (in phase),  $Q=2+2$ product ansatz Hopfions as a
function of the orientation parameters $R$ and $\Phi$ for fixed angles $\Theta=0$ a) and $\Theta=\pi/2$ b).
}\label{fig:9}
\end{figure}

\begin{figure}[hbt]
\begin{center}
a)\hspace{-0.6cm}
\includegraphics[height=.2\textheight,angle =0]{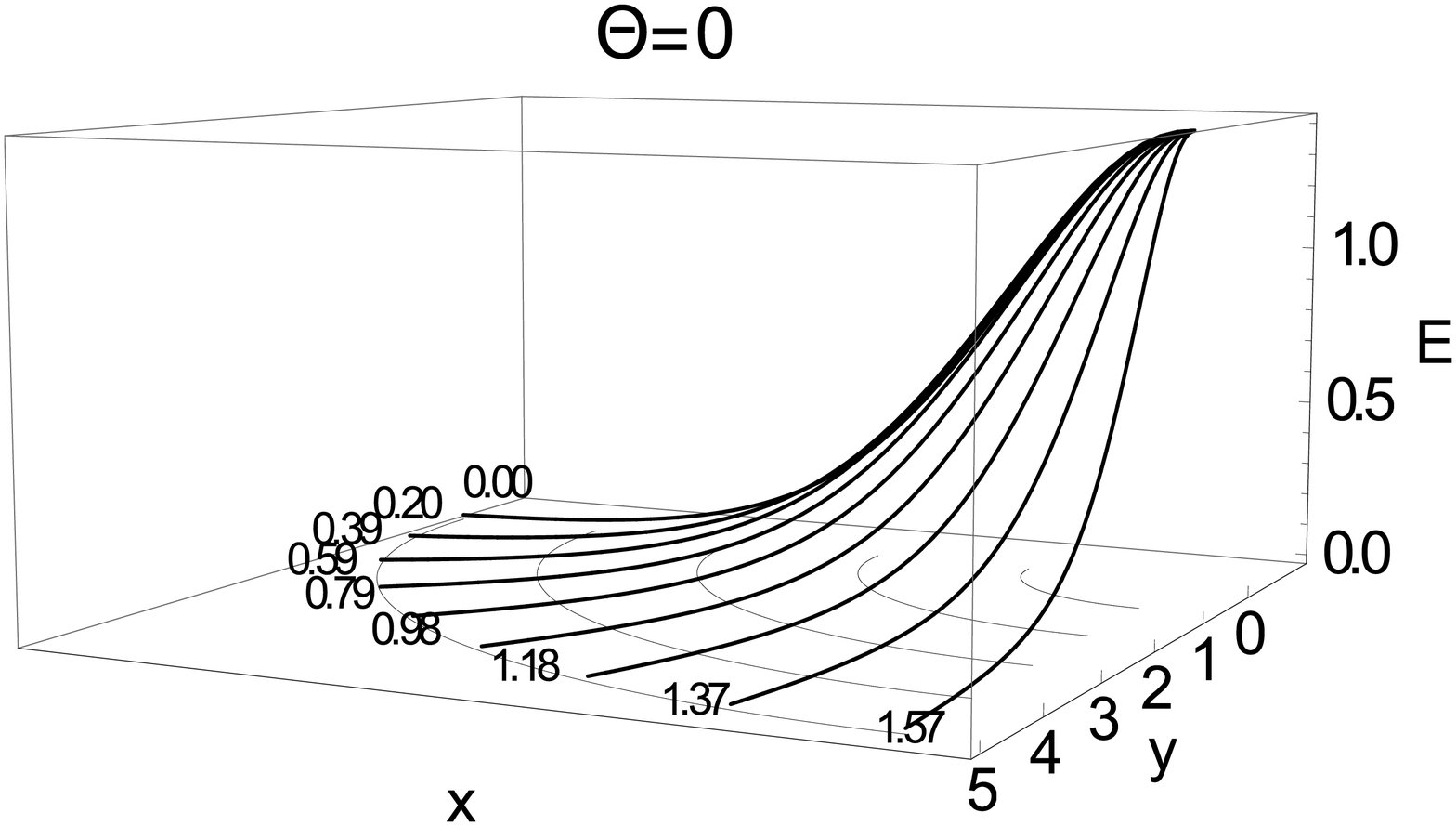}
b)\hspace{-0.6cm}
\includegraphics[height=.2\textheight, angle =0]{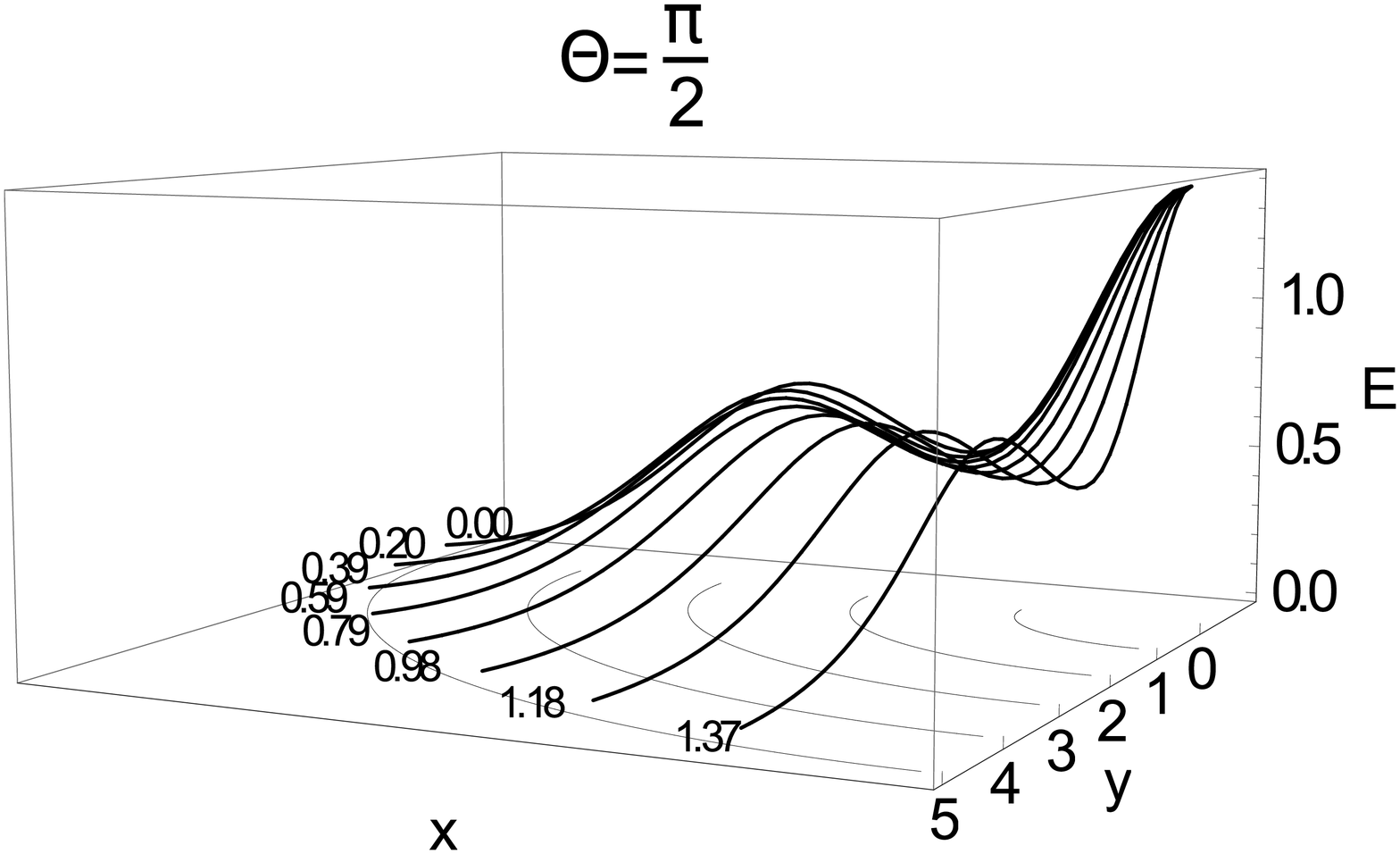}

\end{center}
\vspace{-0.5cm}
\caption{\small The evaluated interaction energy of the $\Delta \alpha =\pi$ (opposite phase), $Q=2+2$ product ansatz Hopfions as a
function of the orientation parameters $R$ and $\Phi$ for fixed angles $\Theta=0$ a) and $\Theta=\pi/2$ b).
}\label{fig:10}
\end{figure}

We have performed calculations with different values of the parameters $R$, $\Theta$ and $\Phi$
for system of two product ansatz $Q=1$ Hopfions~\cite{Acus:2014} (figures \ref{fig:2} and \ref{fig:3}),
system consisting of $Q=1$ and $Q=2$ Hopfions (figures \ref{fig:4} -- \ref{fig:7}), and for
two $Q=2$ Hopfions (figures \ref{fig:8} -- \ref{fig:10}).
All figures demonstrate the integrated interaction energy as a function of
the orientation parameters for the "in phase" Hopfions and the Hopfions with oposite phases.
Clearly, we can expect our approximation of the interaction energy in the
system of two separated Hopfions will be reliable only if the separation parameter $R$
is larger than the sum of cores of the constituents $r_c$.

Our results show that system of two $Q=1$ Hopfions approximated via the product ansatzes
(\ref{hopfionProductAnsatz}) and (\ref{hopfionProductAnsatzAntiParallel})
is in complete agreement with interaction pattern of the Hopfions based on the
simplified dipole-dipole approximation \cite{Ward:2000qj}.
It should be noted that there was a minor mistake in our evaluation of the weight factors of the $L_2$ and $L_4$ terms in the
effective Largangian presented previously in~\cite{Acus:2014}.
Fortunately, this mistake may affect the results only in the case of small values of the
separation parameter $R$, i.e. when the Hopfions cannot be considered as individual constituents, thus
all predictions of~\cite{Acus:2014} remain valid.

In particulary, when the Hopfions are in phase and $\Theta = 0$, which corresponds to Channel A in \cite{Ward:2000qj},
there is a shallow ($-0.05$ at $R\approx 2.6$) attractive window
for separations $R$ large than $1.8$, as can be seen from Fig.~\ref{fig:2}.
Evidently, this attractive channel is very narrow because the potential of interaction
quickly becomes repulsive as the value of $\Theta$ increases.

When Hopfions are in side by side position, $\Theta=\pi/2$,
the interaction potential is always repulsive as is shown in Fig.~\ref{fig:2}.
The interaction strength for other orientations of the Hopfions is represented by a surface in Fig.~\ref{fig:2}.

Quite different pattern of interaction, however, occurs between the opposite phase ${\cal A}_{1,1}$ Hopfions, which is depicted in Fig.~\ref{fig:3}.
In contrast with Fig.~\ref{fig:2} in the Channel A ($\Theta = 0$)
the interaction is always repulsive for all values of the separation distance~$R$.

However, in the Channel B~\cite{Ward:2000qj} ($\Theta = \pi/2$) the interaction energy has relatively
large negative value ($-0.266$ at $R=1.6$) at the separation of about two cores $r_c = 0.8763$ (the interaction is attractive till $R\approx1.1$)
and then gradually decreases to zero as separation
between the Hopfions increases, Fig.~\ref{fig:3}.
The repulsive behaviour changes to attraction at $\Theta \approx \pi/3$, it approaches maximum as $\Theta = \pi/2$.
For the Hopfions which are in opposite phases, this pattern is presented in Fig.~\ref{fig:3} where we plotted the
interaction energy as function of the orientation parameters $R$ and $\Theta$.
Qualitatively, the pattern of interaction between the Hopfions both in the Channel A and in the Channel B,
is in a good agreement with results of full 3d numerical simulations of the Hopfions dynamics~\cite{Hietarinta:2011qk}.

Unfortunately there is no similar 3d~simulation data neither for ${\cal A}_{1,1}$, ${\cal A}_{2,1}$ configurations,
${\cal A}_{1,1}$ and ${\cal A}_{2,1}$ (figures \ref{fig:4} -- \ref{fig:7}), nor for two ${\cal A}_{2,1}$ Hopfions
(figures \ref{fig:8} -- \ref{fig:10}) interactions.
Below we consider the product ansatz approximation of these systems.

It is clear, that in these cases the configuration are less symmetric and the
interaction profiles are more involved, generally they
depend on the value of the azimuthal angle $\Phi$.
In the case of ${\cal A}_{1,1}$ and ${\cal A}_{2,1}$ Hopfion interaction,
which in short will be denoted as $1+2$, the shape of interaction energy isosurface for
fixed values of the angles $\Phi=0$ and $\Phi=\pi/2$  are shown in Figs~\ref{fig:4} (in phase) and \ref{fig:6}
(the case of opposite phases). We see that contrary to $1+1$ case strongest attraction
(interaction energy minimum is $-0.168$ at $R\approx 2.2$) now is observed for Hopfions in phase~(Fig.~\ref{fig:4}),
whereas for the oppositely oriented Hopfions~(cf Fig~\ref{fig:6}) there is only a shallow minimum
($-0.014$ at $R\approx 2.4$) for some narrow interval of values of the orientation angle $\Theta$.

In the case of interaction between two Hopfions ${\cal A}_{2,1}$ ($2+2$ case), strongest attraction channel occurs again for the
Hopfions with opposite phase, similar to the interaction pattern in the $1+1$ case.
The interaction energy minimum, however, becomes shallower with increasing of the Hopfion charges.
In particular, in the $2+2$ case the interaction energy approaches its minimal value  $-0.165$ at $R\approx 1.4$,
compared to $-0.266$ at $R=1.6$ in the $1+1$ case.

In the case of the $1+2$ system the interaction energy, in general, has nontrivial dependency
on the value of the azimuthal angle $\Phi$, as  shown in Fig~\ref{fig:5}.
In a particular case $\Theta=0$
(one Hopfion is above the other), the system possesses the axial symmetry and the
dependency of the interaction potential on the value of $\Phi$ is trivial.

However for other values of polar angle$\Theta$, for example for $\Theta=\pi/2$ (the Hopfions are
in the horizontal plane), this dependency can be evidently seen.
Note that in the cases of the $1+1$ and $2+2$ systems there is no dependency of the interaction energy
on the value of the azimuthal orientation angle $\Phi$.

In our consideration of the interaction between the Hopfions we mainly concentrate ourselves on the
description of possible attractive channels.
Note that all the possible repulsive channels  are illustrated in Figs.~\ref{fig:2}--\ref{fig:10}, thus we
may also draw some conclusions about  the repulsion
strength between the Hopfions for different orientation angles and separation distance $R$.

\section*{Conclusion}

We have investigated various interaction channels in the interaction of the axially symmetric
${\cal A}_{1,1}$ and ${\cal A}_{2,1}$ Hopfions.
The product ansatz of Hopfions can be obtained via coset projection of the corresponding Skyrme field.
This approximation preserves the topological
charge in the entire interaction region. In particular, we analysed
how the interaction energy depends on the orientation
parameters, the separation $R$, the polar angle $\Theta$ and the azimuthal angle $\Phi$.

We have shown that this approach correctly reproduces both the repulsive and attractive
interaction channels discussed previously in the limit of the dipole-dipole interactions for ${\cal A}_{1,1}$.
Using the product ansatz we also were able to predict interaction pattern for pair of ${\cal A}_{1,1}$ and ${\cal A}_{2,1}$,
and two ${\cal A}_{2,1}$ Hopfions. In all cases the interaction has attractive channel
for specific orientations and large
enough separation distances $R$.

Finally, let us note that the product ansatz can be applied to construct a system of interacting Hopfions of even
higher degrees and specific spacial patterns.
An approximation to the higher charge linked solitons, whose position curve consists of a few disjoint loops, like for example
the configuration in the sector of degree $Q = 6$, can be obtained as a multiple product
of the projected matrices~(\ref{hopfion1Projection}).

\acknowledgments
This work is supported by the A.~von Humboldt Foundation (Ya.S.) and also from European
Social Fund under Global Grant measure, VP1-3.1-\v{S}MM-07-K-02-046 (A.A.).

\begin{small}

\end{small}

\end{document}